\documentclass[12pt]{article}
\pdfoutput=1

\usepackage[bulletsep]{collref}
\usepackage{amssymb,graphicx}
\usepackage[intlimits]{amsmath}
\usepackage{bbm}
\usepackage[small]{subfigure}

\usepackage{MnSymbol}
\usepackage{dsfont}
\usepackage[retainorgcmds]{IEEEtrantools}

\makeatletter \@addtoreset{equation}{section} \makeatother

\makeatletter
\let\old@startsection=\@startsection
\let\oldl@section=\l@section
\renewcommand{\@startsection}[6]{\old@startsection{#1}{#2}{#3}{#4}{#5}{#6\mathversion{bold}}}
\renewcommand{\l@section}[2]{\oldl@section{\mathversion{bold}#1}{#2}}
\makeatother

\makeatletter
\let\old@makecaption=\@makecaption
\def\@makecaption{\small\old@makecaption}
\makeatother

\newcommand{\x}{{\tt x}}
\newcommand{\Pexp}{\overrightarrow{\rm P}exp}

\newcommand{\U}{\mathbbm{U}}
\newcommand{\Uh}{\widehat{\mathbbm{U}}}

\newcommand{\K}{\mathcal{K}}
\newcommand{\Q}{\mathcal{Q}}
\newcommand{\eq}{\stackrel{!}{=}}

\usepackage{hyperref}
\hypersetup{pdfstartview={XYZ null null 0.75}}
\begin{document}

%%%%%%%%%%%%%%%%%%%%%%%%%%%%%%%%%%%%%%%%%%%%%%%%%%%%%%%%%%%%%%%%%%%%%%%%%%%%%%%%

\thispagestyle{empty}
\begin{flushright}\footnotesize
%\texttt{ITEP-TH-nn/yy}\\
\texttt{NORDITA 2021-014}
\vspace{0.6cm}
\end{flushright}

\renewcommand{\thefootnote}{\fnsymbol{footnote}}
\setcounter{footnote}{0}

\begin{center}
{\Large\textbf{\mathversion{bold} String integrability of defect CFT \\ and dynamical reflection matrices}
\par}

\vspace{0.6cm}

\textrm{Georgios~Linardopoulos$^{1}$\footnote{Also at the Department of Nuclear and Particle Physics, NKU of Athens, Greece.} and
Konstantin~Zarembo$^{2,3}$\footnote{Also at ITEP, Moscow, Russia.}}
\vspace{8mm}

\textit{${}^1$Institute of Nuclear and Particle Physics, National Centre for Scientific Research, "Demokritos", 153 10, Agia Paraskevi, Greece} \\[6pt]
%\textit{${}^2$Department of Nuclear and Particle Physics, Faculty of Physics, National and Kapodistrian University of Athens, 157 84 Athens, Greece} \\
\textit{${}^2$Nordita, KTH Royal Institute of Technology and Stockholm University, Roslagstullsbacken 23, SE-106 91 Stockholm, Sweden} \\[6pt]
\textit{${}^3$Niels Bohr Institute, Copenhagen University, Blegdamsvej 17, 2100 Copenhagen, Denmark} \\[6pt]
\vspace{0.2cm}
\texttt{glinard@inp.demokritos.gr, zarembo@nordita.org}
%\vspace{3mm}

\vspace{3mm}

%%%%%%%%

\par\vspace{1cm}

\textbf{Abstract} \vspace{3mm}

\begin{minipage}{13cm}
The D3-D5 probe-brane system is holographically dual to a defect CFT which is known to be integrable. The evidence comes mainly from the study of correlation functions at weak coupling. In the present work we shed light on the emergence of integrability on the string theory side. We do so by constructing the double row transfer matrix which is conserved when the appropriate boundary conditions are imposed. The corresponding reflection matrix turns out to be dynamical and depends both on the spectral parameter and the string embedding coordinates.
\end{minipage}

\end{center}

\vspace{0.5cm}

%\newpage
\setcounter{page}{1}
\renewcommand{\thefootnote}{\arabic{footnote}}
\setcounter{footnote}{0}

%%%%%%%%%%%%%%%%%%%%%%%%%%%%%%%%%%%%%%%%%%%%%%%%%%%%%%%%%%%%%%%%%%%%%%%%%%%%%%%%
\flushbottom
\newpage\section[Introduction]{Introduction}
\noindent The string sigma model on AdS$_5 \times \text{S}^5$ \cite{Metsaev:1998it} is an integrable two-dimensional field theory \cite{Bena:2003wd}. Integrability has important implications for the AdS/CFT correspondence extending, via holography, to four dimensions \cite{Minahan:2002ve} and giving rise to powerful tools to explore the duality at the non-perturbative level. \\
\indent The sigma model ordinarily describes closed strings which are automatically integrable. Integrability of a string ending on a D-brane is not so obvious because it can be broken by the boundary conditions at the string's endpoint. The question of which D-branes in AdS$_5 \times \text{S}^5$ are integrable is non-trivial and was addressed in detail in \cite{Dekel:2011ja}. \\
\indent D-branes arise in a variety of holographic setups. An example we will be concerned with is field theory in the presence of a domain wall. The setup is realized by imposing Nahm-pole boundary conditions \cite{Gaiotto:2008sa}, which break the gauge symmetry from $SU(N+k)$ to $SU(N)$ on one side of the domain wall. The system preserves scale invariance and gives rise to a defect CFT (dCFT) \cite{Karch:2000gx,DeWolfe:2001pq,Nagasaki:2011ue,Nagasaki:2012re}. In string theory, the domain wall is a footprint of the D3-D5 intersection. Its holographic dual is a probe D5-brane embedded in AdS$_5$ as a 4-dimensional hyperplane \cite{Karch:2000gx}:
\begin{equation}\label{D5braneEmbedding}
x_3 = \kappa z, \qquad \kappa \equiv \frac{\pi k}{\sqrt{\lambda}} \equiv \tan\alpha,
\end{equation}
where $\lambda$ is the 't~Hooft coupling and the angle $\alpha$ specifies the inclination of the D5-brane relative to the hyperplane $x_3 = 0$. In the standard Poincar\'e coordinates of AdS$_5$,
\begin{equation} \label{PoincareMetric}
ds^2 = \frac{dx_\mu dx^{\mu} + dz^2}{z^2},
\end{equation}
the hyperplane $x_3 = \kappa z$ has the AdS$_4$ geometry. The S$^5$ embedding of the D5-brane is an equatorial S$^2$ with $k$ units of internal gauge field flux:
\begin{equation}\label{FluxinS5}
F = \frac{k}{4} \, \varepsilon _{ijk} \, x^i \, dx^j \wedge dx^k, \qquad i,j,k = 4,5,6,
\end{equation}
where $x_i$ are the 6d Cartesian coordinates describing the embedding ${\rm S}^2\subset{\rm S}^5$. \\
\indent The non-magnetic D5-brane with $k=0$ is known to be integrable \cite{Dekel:2011ja} (see also \cite{DeWolfe:2004zt,Mann:2006rh,Correa:2008av,Correa:2011nz,MacKay:2011zs}), but a D5-brane with an arbitrary inclination angle $\alpha$ and nonzero magnetic flux does not fall into the classification scheme of \cite{Dekel:2011ja}. Yet there is overwhelming evidence that integrability persists for any value of $k$. The evidence comes mainly from the field-theory side where efficient integrability-based techniques have been developed to compute one-point functions of local operators, in perturbation theory \cite{deLeeuw:2015hxa,Buhl-Mortensen:2015gfd,Buhl-Mortensen:2016pxs,deLeeuw:2016umh,Buhl-Mortensen:2016jqo,deLeeuw:2018mkd} and beyond \cite{Buhl-Mortensen:2017ind} (these developments are summarized in \cite{deLeeuw:2017cop,deLeeuw:2019usb,Linardopoulos:2020jck}). Moreover, integrability bootstrap solves for dCFT correlation functions at any coupling \cite{Komatsu:2020sup,Gombor:2020kgu,Gombor:2020auk}. Lack of basic understanding why the D3-D5 system is integrable makes this picture incomplete, we believe. Our goal is to fill this gap. \\
\indent The conserved charges of an integrable system with a boundary are encoded in the double row transfer matrix \cite{Sklyanin:1987bi} which is built from the Lax connection and the reflection matrix. The latter typically has constant numerical entries. Under this assumption, an elegant classification scheme of integrable boundary conditions has been put forward \cite{Dekel:2011ja} establishing a one-to-one link between integrable D-branes and $\mathbbm{Z}_2$ automorphisms of the underlying symmetry algebra. \\
\indent Constancy and independence from the spectral parameter are very natural assumptions. However, there are examples of integrability-preserving boundary conditions, going back to the work of Corrigan and Sheng \cite{Corrigan:1996nt}, that are not described by constant reflection matrices. In principle, the reflection matrix can depend on the spectral parameter, or the dynamical variables, or both. The dynamical reflection matrices arise, for example, in $O(N)$ models with Robin boundary conditions \cite{Aniceto:2017jor,Gombor:2018ppd}. Robin (i.e.\ mixed Neumann-Dirichlet) boundary conditions are precisely the ones that describe a string attached to a D-brane with internal magnetic flux. This explains, perhaps, why the classification of \cite{Dekel:2011ja} missed the magnetized D3-D5 system, and points towards the dynamical character of reflection in this case. \\
\indent Since both AdS$_5$ and S$^5$ are symmetric spaces, we start by briefly reviewing integrability in symmetric cosets with boundaries. We then construct the reflection matrix of a string ending on the D5-brane \eqref{D5braneEmbedding}, \eqref{FluxinS5}. This is done separately in \S\ref{Section:D-braneAdS} for AdS$_5$ and in \S\ref{Section:D-braneS} for S$^5$, because the corresponding equations of motion decouple in the conformal gauge. In section \ref{Section:ConservedCharges} we discuss symmetries and in section \ref{Section:Fermions} we comment on how to include the fermions.
\section[Coset sigma models with boundaries]{Coset sigma models with boundaries \label{Section:CosetSigmaModels}}
A symmetric coset space $G/H_0$ is defined by a $\mathbbm{Z}_2$ decomposition of its symmetry algebra, $\mathfrak{g} = \mathfrak{h}_0 \oplus \mathfrak{h}_2$. The sigma model current gets decomposed as
\begin{equation} \label{MovingFrameCurrent}
J = g^{-1}dg = J_0 + J_2.
\end{equation}
Gauge transformations act on $g \in G$ from the right $g \rightarrow gh_0$, under which $J_0$ transforms as a gauge field and $J_2$ as a matter field in the adjoint. The equations of motion are equivalent to the flatness of the Lax connection \cite{Eichenherr:1979ci}:\footnote{More details can be found in many relevant reviews, e.g.\ \cite{Zarembo:2017muf}.}
\begin{equation}\label{LaxMovingFrame}
L(\x) = J_0 + \frac{\x^2 + 1}{\x^2 - 1}\,J_2 - \frac{2\x}{\x^2 - 1}\, \star J_2 \equiv J + A(\x),
\end{equation}
where $\x$ is the spectral parameter. Defining the fixed frame current,
\begin{equation} \label{FixedFrameCurrent}
j = g J_2 g^{-1},
\end{equation}
the spectral parameter-dependent part $A(\x)$ of the Lax connection \eqref{LaxMovingFrame} takes the following form, in the fixed frame:
\begin{equation}\label{LaxFixedFrame}
a = g A g^{-1} = \frac{2}{\x^2 - 1}\left(j - \x*j\right).
\end{equation}
The connection $a(\x)$ depends only on the matter current $j$. It is also flat:
\begin{equation}
da + a \wedge a = 0.
\end{equation}
\paragraph{Integrable boundary conditions} String dynamics takes place for $\sigma >0$ with some boundary conditions imposed at $\sigma =0$.\footnote{We are considering a semi-infinite string. This is not a real restriction because integrability is broken (or preserved) locally. For example, if
the two ends of the string are attached to the D-brane, integrability imposes two independent conditions at each of the endpoints.} An infinite tower of conserved charges can be constructed by expanding the monodromy matrix
\begin{equation}\label{MonodromyMatrix1}
\mathcal{M}(\x) = g(0){\rm \Pexp} \left(\int_{0}^{\infty}ds \, L_\sigma(s;\x)\right)
\end{equation}
around appropriate values of the spectral parameter $\x$, see \S\ref{Section:ConservedCharges} below. The object \eqref{MonodromyMatrix1} is gauge invariant and its time derivative depends on the matter part of the Lax connection
\begin{equation}\label{MonodromyDerivative}
\dot{\mathcal{M}}(\x) = -a_\tau(0;\x) \mathcal{M}(\x).
\end{equation}
The double row monodromy matrix takes into account the boundary at $\sigma = 0$. It is constructed by folding two monodromy matrices together and connecting them through a reflection matrix:\footnote{We assume that the currents are taken in some representation of the symmetry algebra so that the monodromy matrix is literally a matrix in this representation. ${\mathcal{M}}^t$ then denotes matrix transposition. More abstractly, ${\mathcal{M}}^t$ can be a $\mathbbm{Z}_2$ transformation on $G^C$ induced by an anti-unitary involution of the Lie algebra $\mathfrak{g}$.}
\begin{equation}\label{DoubleRowMonodromyMatrix1}
T(\x) = \mathcal{M}^t(-\x) \U(\x) \mathcal{M}(\x).
\end{equation}
Apart from its explicit dependence on the spectral parameter $\x$, the reflection matrix $\U$ can also be dynamical. In other words, $\U$ may depend on the embedding coordinates at the string's endpoint and through them implicitly on time. The time derivative of the double row monodromy matrix \eqref{DoubleRowMonodromyMatrix1} follows from \eqref{MonodromyDerivative}:
\begin{equation}\label{DoubleRowDerivative}
\dot{T}(\x) = \mathcal{M}^t(-\x) \left(\dot{\U} - a^t_\tau(-\x)\U(\x) - \U(\x)a_\tau(\x)\right) \mathcal{M}(\x).
\end{equation}
If the time derivative \eqref{DoubleRowDerivative} vanishes, the double row transfer matrix will generate infinitely many conserved charges. Therefore the boundary conditions at $\sigma = 0$ preserve integrability if
\begin{equation}\label{IntegrabilityCondition1}
\dot{\U}(\x) \eq a^t_\tau(-\x)\U(\x) + \U(\x)a_\tau(\x),
\end{equation}
where the symbol $\eq$ denotes restriction to the boundary at the string's endpoint $\sigma = 0$, i.e.
\begin{equation}
A \eq B \quad \text{iff} \quad A(\tau,0) = B(\tau,0).
 \end{equation}
Plugging the connection \eqref{LaxFixedFrame} into the integrability condition \eqref{IntegrabilityCondition1} we obtain
\begin{equation}\label{IntegrabilityCondition2}
\dot{\U} \eq \frac{2}{\x^2-1}\Big[\left\{j_\tau^t \, \U + \U j_\tau\right\} + \x\left\{j_\sigma^t \, \U - \U j_\sigma\right\}\Big].
\end{equation}
In the simplest case, the reflection matrix $\U$ is a constant matrix that depends neither on the spectral parameter $\x$ nor on time $\tau$. In this case the integrability condition \eqref{IntegrabilityCondition2} reduces to
\begin{equation}\label{IntegrabilityConditionSimplified}
j_\tau^t \, \U + \U j_\tau \eq j_\sigma^t \, \U - \U j_\sigma \eq 0.
\end{equation}
\section[D-brane in AdS$_5$]{D-brane in AdS$_5$ \label{Section:D-braneAdS}}
The coset AdS$_5 = SO(4,2)/SO(4,1)$ can be realized by means of the 5-dimensional Dirac matrices $\gamma_a$ in the $(-++++)$ signature. The denominator algebra $\mathfrak{so}(4,1)$ is spanned by $\gamma_{ab}$, while $\mathfrak{so}(4,2)$ contains both $\gamma_{ab}$ and $\gamma_a$, i.e.\ $\mathfrak{so}(4,2) = \langle\gamma_a,\gamma_{ab}\rangle$. The coset decomposition is thus
\begin{equation}\label{CosetAdS5}
\mathfrak{h} = \mathfrak{h}_0\oplus\mathfrak{h}_2, \qquad \mathfrak{h}_0 = \left\langle \gamma_{ab} \right\rangle, \qquad \mathfrak{h}_2 = \left\langle \gamma_a \right\rangle,
\end{equation}
where $a,b = 0,\ldots,4$. The Dirac matrices satisfy
\begin{equation}\label{GammaTransposition}
\gamma_a^t = \K^{-1} \gamma_a \K, \qquad \gamma_{ab}^t = -\K^{-1} \gamma_{ab} \K.
\end{equation}

In the chiral representation $\K = \gamma_{13}$, but we will never need this explicit form. Below, we will also split the Dirac matrices into the $SO(3,1)$ components $\gamma_\mu$, $\mu =0 \ldots 3$ and $\gamma_4$, as well as use the chiral projectors
\begin{equation}
\Pi_\pm = \frac{1 \pm \gamma_4}{2}\,.
\end{equation}

The integrability condition \eqref{IntegrabilityCondition2} can be concisely formulated in terms of the transposition brackets, which are defined as
\begin{equation}\label{TranspositionBrackets}
\left\langle A,B\right\rangle_\pm = \K A^t \K^{-1}B \pm BA,
\end{equation}
for any two matrices $A$, $B$. The integrability condition \eqref{IntegrabilityCondition2} is then equivalent to
\begin{equation}\label{IntegrabilityCondition3}
\frac{d\widehat{\U}}{d\tau } \eq \frac{2}{\x^2-1}\left(\left\langle j_\tau,\Uh\right\rangle_+ + \x\left\langle j_\sigma,\Uh\right\rangle_-\right),
\end{equation}
where
\begin{equation}\label{Twist}
\widehat{\U} = \K\U.
\end{equation}

The bracket of the Dirac matrices follows readily from their transposition properties \eqref{GammaTransposition}:
\begin{equation}\label{TranspositionBracketProperties1}
\left\langle\gamma_a,\Gamma\right\rangle_\pm = \left[\gamma_a,\Gamma\right]_\pm, \qquad
\left\langle\gamma_{ab},\Gamma\right\rangle_\pm = -\left[\gamma_{ab},\Gamma\right]_\mp,
\end{equation}
where $\Gamma$ is any matrix. Some useful identities of the transposition brackets are listed in appendix~\ref{Appendix:TranspositionIdentities}. The standard generators of the conformal algebra are
\begin{equation}\label{ConformalGenerators}
D = \frac{\gamma_4}{2}, \qquad P_\mu = \Pi_+ \gamma_\mu, \qquad K_\mu = \Pi_- \gamma_\mu, \qquad L_{\mu\nu} = \gamma_{\mu\nu}.
\end{equation}
\paragraph{Coset representative and current} Choosing the coset representative
\begin{equation}
g = {\rm e}\,^{P_\mu x^\mu }z^{D}, \label{CosetRepresentativeAdS5}
\end{equation}
the $\mathbbm{Z}_2$ decomposition of the moving frame current \eqref{MovingFrameCurrent} takes the form
\begin{equation}
J_0 = \frac{1}{2z} \gamma_{4\mu}dx^\mu, \qquad J_2 = \frac{1}{2z}\left(\gamma_4 dz + \gamma_\mu dx^\mu\right).
\end{equation}
To verify that \eqref{CosetRepresentativeAdS5} correctly describes AdS$_5$, we note that the quadratic form of the Lie algebra
\begin{equation}
\mathop{\mathrm{tr}}\left[J_2^2\right] = \frac{dx_\mu dx^{\mu}+ dz^2}{z^2},
\end{equation}
reproduces the Poincar\'e metric \eqref{PoincareMetric}. The fixed frame current \eqref{FixedFrameCurrent} follows by elementary Dirac algebra:
\begin{equation}\label{FixedFrameCurrentAdS}
j = \frac{1}{2z^2}\left[2\left(zdz + x dx\right)\left(D - x P\right) + \left(z^2 + x^2\right)P dx + K dx + L_{\mu\nu}x^\mu dx^\nu\right],
\end{equation}
where the single summations over covariant indices have been omitted (e.g.\ $x dx \equiv x^\mu dx_\mu$) and $x^2 \equiv x_\mu x^\mu$.
\paragraph{Boundary conditions} The longitudinal and transverse coordinates of the tilted AdS$_4$ brane \eqref{D5braneEmbedding} are:
\begin{IEEEeqnarray}{rll}
\text{longitudinal}:\quad & x_{0,1,2}, \quad & x_\parallel \equiv x_3\sin\alpha + z\cos\alpha \qquad \label{LongitudinalCoordinatesAdS}\\
\text{transverse}:\quad && x_\perp \equiv x_3 \cos\alpha - z\sin\alpha. \qquad \label{TransverseCoordinatesAdS}
\end{IEEEeqnarray}
The string boundary conditions on the D5-brane are Neumann for the longitudinal coordinates ($x_{0,1,2}$, $x_\parallel$) and Dirichlet for the transverse coordinate ($x_\perp$), i.e.\
\begin{IEEEeqnarray}{rl}
\acute{x}_{0,1,2} \eq \acute{x}_\parallel \eq 0 \qquad & (\text{Neumann}) \label{BoundaryConditionsAdS1} \\
\dot{x}_\perp \eq x_\perp \eq 0 \qquad & (\text{Dirichlet}). \label{BoundaryConditionsAdS2}
\end{IEEEeqnarray}
Below we determine a reflection matrix $\Uh$ that satisfies the integrability condition \eqref{IntegrabilityCondition3} upon imposing \eqref{BoundaryConditionsAdS1}--\eqref{BoundaryConditionsAdS2}, thus showing that the D5-brane is integrable in AdS$_5$.
\subsection[Vertical brane]{Vertical brane}
Let us first consider the non-magnetic D5-brane for which $k = 0$. \eqref{D5braneEmbedding} then implies that the inclination of the brane relative to the hyperplane $x_3 = 0$ is zero, i.e.\ $\alpha = 0$. The brane is perpendicular to the $x_3$ axis. At the string's endpoint,
\begin{eqnarray}
j_\tau &\eq& \frac{1}{2z^2}\left[2\left(z\dot{z} + x^i\dot{x}_i\right)(D - x^j P_j) + \left(z^2 + x_i^2\right) \dot{x}^j P_j + \dot{x}^i K_i + x^i \dot{x}^j L_{ij}\right] \qquad \label{FixedFrameCurrentAdStau} \\
j_\sigma &\eq& \frac{\acute{x}_3}{2z^2}\left[\left(z^2 + x_i^2\right)P_3 + K_3 + x^i L_{i3}\right], \qquad \label{FixedFrameCurrentAdSsigma}
\end{eqnarray}
where $i,j = 0,1,2$. Note that the conformal generators \eqref{ConformalGenerators} making up the fixed frame current \eqref{FixedFrameCurrentAdS} split into two disjoint groups, those that appear in $j_\tau$ and those that appear in $j_\sigma$. The two equations in \eqref{IntegrabilityConditionSimplified} therefore decouple and can be solved separately. \\
\indent In terms of the transposition brackets \eqref{TranspositionBrackets}, the integrability condition \eqref{IntegrabilityConditionSimplified} is equivalent to a set of seven equations:
\begin{IEEEeqnarray}{rrrrl}
\left\langle D,\widehat{\U}\right\rangle_+ = 0,& \qquad \left\langle P_i,\Uh\right\rangle_+ = 0,& \qquad \left\langle K_i,\Uh\right\rangle_+ = 0,& \qquad \left\langle L_{ij},\Uh\right\rangle_+ = 0& \qquad \label{IntegrabilityConditionVertical1}\\
& \left\langle P_3,\Uh\right\rangle_- = 0,& \qquad \left\langle K_3,\Uh\right\rangle_- = 0,& \qquad \left\langle L_{i3},\Uh\right\rangle_- = 0&. \qquad \label{IntegrabilityConditionVertical2}
\end{IEEEeqnarray}
Given that the brane is perpendicular to the $x_3$ axis, its normal four-vector can be written as $n_{\mu} = (0,0,0,1)$, so that the most natural solution of \eqref{IntegrabilityConditionVertical1}--\eqref{IntegrabilityConditionVertical2} is
\begin{equation}\label{ReflectionMatrixVerticalBraneAdS}
\Uh = n^{\mu}\gamma_{\mu} = \gamma_3.
\end{equation}
Indeed, by using \eqref{TranspositionBracketProperties1}, \eqref{TranspositionBracketProperties3} one can readily check that \eqref{ReflectionMatrixVerticalBraneAdS} satisfies \eqref{IntegrabilityConditionVertical1}--\eqref{IntegrabilityConditionVertical2} and thus \eqref{IntegrabilityCondition3} holds for the constant reflection matrix, in accordance with the findings of \cite{Dekel:2011ja}.
\subsection[Inclined brane]{Inclined brane}
We now consider an arbitrary inclination angle $\alpha$. Introducing the longitudinal coordinates,
\begin{equation}
y_\mu = \left(x_i, x_\parallel\right),
\end{equation}
the string boundary conditions \eqref{BoundaryConditionsAdS1}--\eqref{BoundaryConditionsAdS2} can be written as
\begin{equation}
\acute{y}_\mu \eq \dot{x}_\perp \eq x_\perp \eq 0.
\end{equation}
We will use the longitudinal coordinates $y_\mu$ alongside the spacetime four-vector $x_\mu =(x_i, x_3)$. Noting the identity
\begin{equation}
x^2 + z^2 = y^2 + x_\perp^2,
\end{equation}
the values of the fixed frame currents \eqref{FixedFrameCurrentAdS} on the boundary (at $x_\perp = 0$) become:
\begin{eqnarray}
j_\tau &\eq& \frac{1}{2z^2}\left[2 y_{\mu} \dot{y}^{\mu} (D - x^{\nu} P_{\nu}) + y^2\dot{x}^{\mu} P_{\mu} + \dot{x}^{\mu} K_{\mu} + x^\mu \dot{x}^\nu L_{\mu\nu}\right] \\
j_\sigma &\eq& \frac{\acute{x}_\perp\cos\alpha}{2z^2} \left(y^2 P_3 + K_3 + x^\mu L_{\mu3}\right).
\end{eqnarray}
\indent Because the equation $\left\langle j_\tau ,\gamma_3\right\rangle_+\eq 0$ no longer holds for the inclined brane, the reflection matrix has to be deformed. The existence of such a deformation, highly non-trivial in itself, crucially depends on the fact that $\left\langle j_\tau ,\gamma_3\right\rangle_+$ is a total derivative:
\begin{equation}\label{magichappens}
\left\langle j_\tau ,\gamma_3\right\rangle_+\eq \kappa \,\frac{dS}{d\tau }, \qquad S \equiv \frac{x^{\mu}\gamma_{\mu} - \Pi_+ - y^2\Pi_-}{z}\,.
\end{equation}
The matrix $S$ satisfies two remarkable identities, which hold once $x_\perp$ is set to zero:
\begin{equation}
\left\langle j_\tau, S\right\rangle_+ \eq -\dot{S}, \qquad \left\langle j_\sigma, S\right\rangle_- \eq 0.
\end{equation}
These can be checked by using \eqref{TranspositionBracketProperties1} and the formulae in appendix~\ref{Appendix:TranspositionIdentities}. These identities suggest the following ansatz for the reflection matrix $\Uh$:
\begin{equation}
\Uh = \gamma_3 + CS,
\end{equation}
where $C$ is a constant that may depend on the spectral parameter. The ansatz goes through the integrability condition \eqref{IntegrabilityCondition3} by virtue of \eqref{magichappens}, leaving behind an algebraic equation for $C$:
\begin{equation}
C = \frac{2}{\x^2 - 1}\left(\kappa - C\right) \Rightarrow C = \frac{2\kappa}{\x^2 + 1}.
\end{equation}
This leads to the following solution for the reflection matrix:
\begin{equation} \label{ReflectionMatrixAdS}
\Uh = \gamma_3 + \frac{2\kappa}{\x^2 + 1}\,\,\frac{x^{\mu} \gamma_{\mu} - \Pi_+ - (x^2 + z^2)\Pi_-}{z}.
\end{equation}
The reflection matrix \eqref{ReflectionMatrixAdS} is dynamical (i.e.\ it depends on the embedding coordinates of the string) and carries a non-trivial dependence on the spectral parameter $\x$. According to the general analysis of coset models with boundaries \cite{Gombor:2018ppd,Gombor:2019bun}, the reflection matrix can be polynomial in the spectral parameter of at most degree two. This is also true for our solution, because
multiplication with $\x^2 + 1$ makes it a quadratic polynomial in $\x$.
\section[D-brane in S$^5$]{D-brane in S$^5$ \label{Section:D-braneS}}
The 5-dimensional Dirac matrices $\gamma_{\grave{a}}$ in the $(+++++)$ signature, represent the coset S$^5 = SO(6)/SO(5)$. The numerator algebra $\mathfrak{so}(6)$ is formed by $\gamma_{\grave{a}}$ and their commutators $\gamma_{\grave{a}\grave{b}}$, while the denominator algebra $\mathfrak{so}(5)$ is spanned by $\gamma_{\grave{a}\grave{b}}$. The coset decomposition reads
\begin{equation}\label{CosetS5}
\mathfrak{h} = \mathfrak{h}_0\oplus\mathfrak{h}_2, \qquad \mathfrak{h}_0 = \left\langle \gamma_{\grave{a}\grave{b}} \right\rangle, \qquad \mathfrak{h}_2 = \left\langle \gamma_{\grave{a}} \right\rangle,
\end{equation}
where $\grave{a},\grave{b} = 1,\ldots,5$. As before, the matrices $\gamma_{\grave{a}}$, $\gamma_{\grave{a}\grave{b}}$ satisfy \eqref{GammaTransposition}, \eqref{TranspositionBracketProperties1}.
\paragraph{Coset representative and current} The coset parametrization of S$^5$ is
\begin{eqnarray} \label{CosetRepresentativeS5}
g = n_6 + i\gamma_{\grave{a}}n_{\grave{a}},
\end{eqnarray}
where
\begin{eqnarray}
n_6 = \cos\frac{\theta}{2}, \qquad n_{\grave{a}} = m_{\grave{a}} \, \sin\frac{\theta}{2}, \qquad m_{\grave{a}}m_{\grave{a}} = 1.
\end{eqnarray}
The coset variables $n_{\grave{a}}$, $n_6$ are quite distinct from the S$^5$ coordinates $x_a$, $x_9$ ($a = 4,\ldots,8$). Choosing the standard S$^5$ parametrization,
\begin{eqnarray}
x_{a} = m_{(a-3)} \, \sin\theta, \qquad x_9 = \cos\theta,
\end{eqnarray}
we obtain the map
\begin{eqnarray}
x_{a} = 2 n_6 \, n_{(a-3)}, \qquad x_9 = 2n_6^2 - 1.
\end{eqnarray}
The $\mathds{Z}_2$ components of the moving frame current \eqref{MovingFrameCurrent} that follow from the coset representative \eqref{CosetRepresentativeS5} are
\begin{eqnarray}\label{MovingFrameCurrentS}
J_0 = \gamma_{\grave{a}\grave{b}}n_{\grave{a}} dn_{\grave{b}}, \qquad J_2 = i\gamma_{\grave{a}}\left(n_{6}dn_{\grave{a}} - n_{\grave{a}}dn_6\right).
\end{eqnarray}
As a crosscheck, the quadratic form
\begin{equation}
-\mathop{\mathrm{tr}}\left[J_2^2\right] = d\theta^2 + \sin^2\theta\left(dm_{\grave{a}}dm_{\grave{a}}\right) = \sum_{a = 4}^9 dx_a^2,
\end{equation}
correctly reproduces the S$^5$ metric. The overall minus sign is related to the fact that the currents \eqref{MovingFrameCurrentS} of the 5-sphere occupy a block in the matrices of the AdS$_5\times\text{S}^5$ supercurrents, the supertrace of which reproduces the metric of the full space. The fixed frame current \eqref{FixedFrameCurrent} is given, in the case of S$^5$, by
\begin{eqnarray} \label{FixedFrameCurrentS}
j = i(2n_6^2 - 1) n_6 dn_{\grave{a}} \gamma_{\grave{a}} - i(2n_6^2 + 1) dn_6 n_{\grave{a}} \gamma_{\grave{a}} - 2n_6^2 n_{\grave{a}}dn_{\grave{b}} \gamma_{\grave{a}\grave{b}}.
\end{eqnarray}
\paragraph{Boundary conditions} The longitudinal and transverse coordinates of the S$^2 \subset \text{S}^5$ component of the D5-brane are
\begin{IEEEeqnarray}{rl}
\text{longitudinal}:\quad & \textbf{x}_{\parallel} = \left(x_4, x_5, x_6\right) \qquad \label{LongitudinalCoordinatesS}\\
\text{transverse}:\quad & \textbf{x}_{\perp} = \left(x_7, x_8, x_9\right). \qquad \label{TransverseCoordinatesS}
\end{IEEEeqnarray}
The string boundary conditions on the 2-sphere are Dirichlet for the transverse coordinates $\textbf{x}_{\perp}$ and, due to the internal flux \eqref{FluxinS5}, Neumann-Dirichlet for the longitudinal coordinates $\textbf{x}_{\parallel}$:
\begin{eqnarray}
\acute{x}_4 &\eq &\kappa \left(x_5 \, \dot{x}_6 - x_6 \, \dot{x}_5\right) \label{BoundaryConditionsS1} \\
\acute{x}_5 &\eq &\kappa \left(x_6 \, \dot{x}_4 - x_4 \, \dot{x}_6\right), \qquad \dot{x}_7 \eq \dot{x}_8 \eq \dot{x}_9 \eq 0 \label{BoundaryConditionsS2} \\
\acute{x}_6 &\eq & \kappa \left(x_4 \, \dot{x}_5 - x_5 \, \dot{x}_4\right). \label{BoundaryConditionsS3}
\end{eqnarray}
In compact form these boundary conditions read
\begin{eqnarray}
\acute{\textbf{x}}_{\parallel} - \kappa \left(\textbf{x}_{\parallel} \times \dot{\textbf{x}}_{\parallel}\right) & \eq & 0 \qquad \text{(Neumann-Dirichlet)} \label{BoundaryConditionsS4} \\
\dot{\textbf{x}}_{\perp} &\eq & 0 \qquad \text{(Dirichlet)}. \label{BoundaryConditionsS5}
\end{eqnarray}
The 5-sphere coordinates $x_a$, $x_9$ also obey,
\begin{eqnarray} \label{BoundaryConditionsS6}
\sum_{a=4}^{6} x_a^2 \eq 1, \qquad x_7 \eq x_8 \eq x_9 \eq 0.
\end{eqnarray}
In terms of the coset variables $n_{\grave{a}}$, $n_6$, the string boundary conditions \eqref{BoundaryConditionsS1}--\eqref{BoundaryConditionsS3} become
\begin{eqnarray}
&n_1\acute{n}_6 + n_6\acute{n}_1 &\eq 2\kappa n_6^2 \left(n_2 \, \dot{n}_3 - n_3 \, \dot{n}_2\right) \label{BoundaryConditionsS7} \\
&n_2\acute{n}_6 + n_6\acute{n}_2 &\eq 2\kappa n_6^2 \left(n_3 \, \dot{n}_1 - n_1 \, \dot{n}_3\right) \qquad \dot{n}_4 \eq \dot{n}_5 \eq \dot{n}_6 \eq 0 \label{BoundaryConditionsS8} \\
&n_3\acute{n}_6 + n_6\acute{n}_3 &\eq 2\kappa n_6^2 \left(n_1 \, \dot{n}_2 - n_2 \, \dot{n}_1\right), \label{BoundaryConditionsS9}
\end{eqnarray}
while also
\begin{eqnarray} \label{BoundaryConditionsS10}
n_1^2 + n_2^2 + n_3^2 \eq \frac{1}{2}, \qquad n_4 \eq n_5 \eq 0, \qquad n_6 \eq \frac{1}{\sqrt{2}}.
\end{eqnarray}
\paragraph{Integrable boundary conditions} To show that the D5-brane is integrable in S$^5$, we need to specify a reflection matrix $\Uh$ which satisfies the integrability condition \eqref{IntegrabilityCondition3} upon imposing the boundary conditions \eqref{BoundaryConditionsS1}--\eqref{BoundaryConditionsS6}, or equivalently \eqref{BoundaryConditionsS7}--\eqref{BoundaryConditionsS10}. The values of the fixed frame currents \eqref{FixedFrameCurrentS} on the $\textbf{x}_{\perp} = 0$ boundary at $\sigma = 0$ become
\begin{IEEEeqnarray}{l}
j_{\tau} \eq - n_{\grave{i}}\dot{n}_{\grave{j}} \gamma_{\grave{i}\grave{j}} \label{FixedFrameCurrentStau} \\
j_{\sigma} \eq - 2i \acute{n}_6 n_{\grave{i}} \gamma_{\grave{i}} - n_{\grave{i}}\left(\acute{n}_{4} \gamma_{\grave{i}4} + \acute{n}_{5} \gamma_{\grave{i}5}\right) - \sqrt{2} \, \kappa \, n_{\grave{i}} n_{\grave{j}} \dot{n}_{\grave{k}} \, \epsilon_{\grave{j}\grave{k}\grave{\ell}} \, \gamma_{\grave{i}\grave{\ell}}. \label{FixedFrameCurrentSsigma}
\end{IEEEeqnarray}
Plugging \eqref{FixedFrameCurrentStau}--\eqref{FixedFrameCurrentSsigma} into the integrability condition \eqref{IntegrabilityCondition3}, one can prove that the reflection matrix
\begin{eqnarray} \label{ReflectionMatrixS}
\Uh = \gamma_{45} + \frac{2\kappa \, \x}{\x^2 + 1} \frac{n_{\grave{i}}\gamma_{\grave{i}}}{n_6},
\end{eqnarray}
where $\grave{i} = 1,2,3$, satisfies it. Therefore the string boundary conditions \eqref{BoundaryConditionsS1}--\eqref{BoundaryConditionsS10} on the 5-sphere are integrable.
\section[Conserved charges]{Conserved charges \label{Section:ConservedCharges}}
As we have already mentioned, the Taylor expansions of the monodromy matrices \eqref{MonodromyMatrix1}, \eqref{DoubleRowMonodromyMatrix1} lead to infinite sets of conserved charges. The expansion at $\x = \infty $ in particular, generates the conserved charges of the global symmetry. The aim of the present section is to determine the set of global symmetries of the string sigma model on AdS$_5\times\text{S}^5$ that is preserved by the D5-brane. \\
\indent We first note that the monodromy matrix \eqref{MonodromyMatrix1} can be written as:\footnote{Assuming appropriate boundary conditions at $\sigma =\infty $.}
\begin{equation}\label{MonodromyMatrix2}
\mathcal{M}(\x) = g(0) \, {\rm\overrightarrow{\rm P}exp}\left(\int_{0}^{\infty }ds\,L_\sigma (s;\x)\right) = {\rm\overrightarrow{\rm P}exp} \left(\int_{0}^{\infty }ds\,a_\sigma (s;\x)\right) ,
\end{equation}
where $a_{\sigma}$ is the $\sigma$-component of the fixed frame Lax connection \eqref{LaxFixedFrame}
\begin{equation}
a_{\sigma}(\x) = \frac{2}{\x^2-1}\left(j_{\sigma} - \x \, j_{\tau}\right).
\end{equation}
Taylor-expanding the path-ordered exponential \eqref{MonodromyMatrix2} around $\x = \infty$ leads to
\begin{equation} \label{MonodromyExpansion}
{\rm\overrightarrow{\rm P}exp} \left(\int_{0}^{\infty }ds\,a_\sigma\right) = \mathbbm{1} - \frac{2}{\x}\int_0^{\infty} ds j_{\tau} + \frac{2}{\x^2} \left[\int_0^{\infty} ds j_{\sigma} + 2\int_0^{\infty}\int_0^{s} ds ds' j_{\tau}^{'} j_{\tau}\right] - \ldots
%
%- \frac{2}{\x^3}\bigg(\int_0^{\infty} j_{\tau}(s)ds + 2 \int_0^{\infty}\int_0^{s_1} \Big(j_{\sigma}(s_2) j_{\tau}(s_1) + j_{\tau}(s_2) j_{\sigma}(s_1)\Big) ds_2 ds_1 + 4\int_0^{\infty}\int_0^{s_1}\int_0^{s_2} j_{\tau}(s_3) j_{\tau}(s_2) j_{\tau}(s_1) ds_3 ds_2 ds_1\bigg) + \ldots
\end{equation}
The expansion \eqref{MonodromyExpansion} gives rise to an infinite tower of (generally nonlocal) conserved charges. To identify these charges we use
\begin{equation}
{\rm\overrightarrow{\rm P}exp} \left(\int_{0}^{\infty }ds\,a_\sigma\right) = \exp\left(2\sum_{r = 0}^{\infty} \left(-\frac{1}{\x}\right)^{r+1} Q_{r}\right) = \mathbbm{1} - \frac{2}{\x}Q_{0} + \frac{2}{\x^2} \left(Q_{1} + Q_{0}^2\right) - \ldots
\end{equation}
The first charge in the above hierarchy is just the Noether charge of the global bosonic symmetry $SO\left(4,2\right) \times SO(6)$ of the string sigma model on AdS$_5\times\text{S}^5$:
\begin{equation}\label{NoetherCharge}
Q_{0} = \int_0^{\infty} ds j_{\tau}.
\end{equation}
\paragraph{Double row monodromy matrix} It may seem that the double row monodromy matrix \eqref{DoubleRowMonodromyMatrix1} generates the same number of conserved charges as the monodromy matrix \eqref{MonodromyMatrix2}, leading to the wrong conclusion that boundaries do not break any symmetries. In practice however, some charges get cancelled by folding (i.e.\ through the construction \eqref{DoubleRowMonodromyMatrix1}) and are simply not there in systems with boundaries. In more detail, we need to expand the monodromy matrix
\begin{equation} \label{DoubleRowMonodromyMatrix2}
T(\x) = {\rm\overleftarrow{\rm P}exp} \left(\int_{0}^{\infty }ds\,a_\sigma^t(s;-\x)\right) \U(\x) {\rm\overrightarrow{\rm P}exp} \left(\int_{0}^{\infty} ds\,a_\sigma(s;\x)\right)
\end{equation}
in $1/\x$. Taking into account the general form of the AdS$_5\times\text{S}^5$ reflection matrices \eqref{ReflectionMatrixAdS}, \eqref{ReflectionMatrixS},
\begin{equation}
\Uh(\x) = \Uh_0 + \frac{1}{\x^2 + 1}\left(\x \, \Uh_1 + \Uh_2\right),
\end{equation}
as well as the expansion \eqref{MonodromyExpansion}, we get the expansion of the double row monodromy matrix $\hat{T} \equiv \K \, T$ around $\x = \infty$:
\begin{equation} \label{DoubleRowMonodromyExpansion}
\hat{T}(\x) = \Uh_0 + \frac{1}{\x}\left(\Uh_1 + 2\int_0^{\infty}ds \left\langle j_{\tau}, \Uh_0\right\rangle_- \right) + \ldots
\end{equation}
In order to identify the conserved charges we set
\begin{equation}
\hat{T}(\x) = \Uh_0 + \frac{2}{\x}\tilde{Q}_{0} + \frac{2}{\x^2} \left(\tilde{Q}_{1} + \tilde{Q}_{0}^2\right) + \ldots,
\end{equation}
finding in particular, for the first conserved charge
\begin{equation} \label{BoundaryCharge}
\tilde{Q}_{0} = \frac{\Uh_1}{2} + \int_0^{\infty}ds \left\langle j_{\tau}, \Uh_0\right\rangle_-.
\end{equation}
\indent By comparing the bulk Noether charge \eqref{NoetherCharge} with the first conserved charge \eqref{BoundaryCharge} on the boundary, we can determine the fraction of the global bosonic symmetry $SO\left(4,2\right) \times SO(6)$ of the AdS$_5\times\text{S}^5$ string sigma model that is preserved by the D5-brane. The preserved symmetries correspond to the set of generators for which the transposition bracket $\left\langle j_{\tau}, \Uh_0\right\rangle_-$ is nonzero. The respective charges are not eliminated by folding. \\
\indent Interestingly, the symmetries that are preserved by the D5-brane are determined by $\Uh_0$ and are thus independent of $k$. The conserved charges on $S^5$ receive an extra contribution from $\Uh_1$ that is localized on the brane, implying that the endpoint of the string carries an R-charge. \\
\indent For AdS$_5$, \eqref{ReflectionMatrixAdS} gives $\Uh_0 = \gamma_3$, $\Uh_1 = 0$, while the transposition identities \eqref{IntegrabilityConditionVertical2} imply
\begin{equation}
\left\langle P_3,\Uh_0\right\rangle_- = \left\langle K_3,\Uh_0\right\rangle_- = \left\langle L_{i3},\Uh_0\right\rangle_- = 0,
\end{equation}
for $i,j = 0,1,2$. These are the broken conformal generators. The preserved AdS$_5$ symmetries are generated by
\begin{equation} \label{PreservedChargesAdS}
\left\{D, P_i, K_i, L_{ij}\right\},
\end{equation}
which spans the $SO(3,2)$ subgroup of $SO(4,2)$. The domain wall preserves the group of 3-dimensional conformal transformations, in agreement with the AdS$_4$ geometry of the D5-brane in AdS. \\
\indent For S$^5$, the reflection matrix \eqref{ReflectionMatrixS} gives $\Uh_0 = \gamma_{45}$, $\Uh_1 = 2\kappa n_{\grave{i}}\gamma_{\grave{i}}/n_6$, whereas
\begin{equation}
\left\langle \gamma_{\grave{i}},\Uh_0\right\rangle_- = \left\langle \gamma_{\grave{i}4},\Uh_0\right\rangle_- = \left\langle \gamma_{\grave{i}5},\Uh_0\right\rangle_- = 0,
\end{equation}
for $\grave{i} = 1,2,3$. Therefore the preserved S$^5$ symmetry consists of the generators
\begin{equation} \label{PreservedChargesS}
\left\{\gamma_{\grave{i}\grave{j}}, \gamma_{4}, \gamma_{5}, \gamma_{45}\right\}.
\end{equation}
The Dirac matrices $\left\{\gamma_{4}, \gamma_{5}, \gamma_{45}\right\}$ satisfy the $\mathfrak{so}(3)$ algebra, and so do $\gamma_{\grave{i}\grave{j}}$. The unbroken symmetry group is thus $SO(3)\times SO(3)$ which again agrees with the D5-brane geometry in S$^5$. The boundary itself carries an R-charge which is proportional to $\Uh_1$ and belongs to the broken part of the symmetry algebra.
\section[Including the fermions]{Including the fermions \label{Section:Fermions}}
\noindent Including the fermions is rather straightforward \cite{Dekel:2011ja}. The symmetry algebra of AdS$_5\times \text{S}^5$ is embedded in $\mathfrak{psu}(2,2|4)$, the Dirac-matrix representation being best suited for this purpose, the $\mathbbm{Z}_2$ symmetry is replaced by $\mathbbm{Z}_4$ \cite{Berkovits:1999zq} and transposition by supertransposition. The reflection matrix is just the direct sum of the AdS$_5$ and S$^5$ components \eqref{ReflectionMatrixAdS}, \eqref{ReflectionMatrixS} that were computed above:
\begin{equation}
U = \begin{bmatrix} \U & 0 \\ 0 & \U' \end{bmatrix},
\end{equation}
where the $\U$ block of the matrix corresponds to AdS, and $\U'$ to the sphere. \\
\indent The Lax connection is built from the $\mathbbm{Z}_4$ components of the current
\begin{equation}
J = g^{-1}dg = J_0 + J_1 + J_2 + J_3,
\end{equation}
where $g$ is now the group element of $PSU(2,2|4)$, the currents $J_0$ and $J_2$ are bosonic (or even), while $J_1$, $J_3$ are fermionic (or odd). The Lax connection reads \cite{Bena:2003wd}
\begin{equation}\label{LaxSuperstring}
L(\x) = J_0 + \frac{\x^2 + 1}{\x^2 - 1}\,J_2 - \frac{2\x}{\x^2 - 1}\,*J_2 + \sqrt{\frac{\x + 1}{\x - 1}}\,J_1 + \sqrt{\frac{\x - 1}{\x + 1}}\,J_3.
\end{equation}
The flatness of $L(\x)$ is equivalent to the full set of equations of motion that follows from the AdS$_5\times\text{S}^5$ superstring action. The monodromy matrix is again given by \eqref{MonodromyMatrix1}, where the Lax connection \eqref{LaxMovingFrame} is replaced by \eqref{LaxSuperstring}. In the double row construction \eqref{DoubleRowMonodromyMatrix1}, transposition gets replaced by supertransposition, for consistency with Grassmann grading:
\begin{equation}
T(\x) = \mathcal{M}^{st}(-\x) U(\x) \mathcal{M}(\x).
\end{equation}
In complete analogy with the bosonic case that was treated in the previous section, the supercharges $Q$ that are broken by the string boundary conditions are determined from the condition:
\begin{equation}\label{brokenQs}
\left\langle Q, \widehat{U}_0 \right\rangle_- = 0,
\end{equation}
where $\left\langle\cdot , \cdot \right\rangle$ now denotes the supertransposition bracket
\begin{equation}
\left\langle A, B\right\rangle_\pm = K A^{st} K^{-1}B \pm BA,
\end{equation}
that generalizes \eqref{TranspositionBrackets}. Moreover, we have defined
\begin{equation}
K = \begin{bmatrix} \K & 0 \\ 0 & \K \\ \end{bmatrix}, \qquad \widehat{U}_0 = \begin{bmatrix} \gamma_3 & 0 \\ 0 & \gamma_{45} \end{bmatrix}.
\end{equation}
Given that the odd elements of $\mathfrak{psu}(2,2|4)$ are of the form
\begin{equation}
Q = \begin{bmatrix} 0 & \Q \\ -\Q^\dagger \gamma_5 & \end{bmatrix}, \qquad Q^{st} = \begin{bmatrix} 0 & \K{}^{-1}\gamma_5\K\Q^* \\ \Q^t & 0 \end{bmatrix},
\end{equation}
it follows from \eqref{brokenQs} that the broken supercharges obey the reality condition,
\begin{equation} \label{SuperchargeRealityCondition}
\Q^* = \K^{-1} \gamma_{35} \Q \gamma_{45}\K.
\end{equation}
The reality condition \eqref{SuperchargeRealityCondition} singles out exactly half of the supercharges which are broken by the boundary conditions. The other half remains unbroken. We conclude that the brane \eqref{D5braneEmbedding}, \eqref{FluxinS5} is one-half BPS, in accordance with the supergravity analysis of \cite{Skenderis:2002vf}.
\section{Conclusions}
\noindent The reflection matrix that defines the hierarchy of conserved charges of a string ending on a D5-brane is maximally complicated since it depends, not only on the spectral parameter, but also on dynamical variables. In quantum theory the D5-brane carries internal degrees of freedom \cite{Komatsu:2020sup}, since the elementary excitations of the string form bound states upon reflection from the boundary. There are $k$ such bound states \cite{Komatsu:2020sup}. The same parameter $k$ controls both the inclination of the brane in AdS$_5$ and the magnetic flux in S$^5$. In the classical regime of string theory the parameter $k$ is very large, scaling naturally as $k\sim\sqrt{\lambda}$. We believe that the proliferation of bound states and the dynamical character of the reflection matrix are not unrelated. Another indication that some degrees of freedom localize on the brane is the boundary contribution to the R-charge that appears in \eqref{BoundaryCharge}. \\
\indent There are other classes of integrable boundary conditions of the string in AdS$_5\times\text{S}^5$ that describe a variety of states and operators in the dual gauge theory. Those associated with constant reflection matrices are completely classified \cite{Dekel:2011ja}. It would be interesting to see which of them admit deformations with dynamical reflection matrices. Extending the analogy with the D5-brane, we expect the deformation parameter to be quantized at finite coupling and to correspond to the dimension of the boundary Hilbert space.
\subsection*{Acknowledgements}
We would like to thank M.\ de Leeuw and C.\ Kristjansen for interesting discussions. G.L.\ received funding from the Hellenic Foundation for Research and Innovation (HFRI) and the General Secretariat for Research and Technology (GSRT), in the framework of the \textit{first post-doctoral researchers support}, under grant agreement No.\ 2595. The work of K.Z.\ was supported by the grant "Exact Results in Gauge and String Theories" from the Knut and Alice Wallenberg foundation and by RFBR grant 18-01-00460 A.
\appendix\section[Transposition bracket identities]{Transposition bracket identities \label{Appendix:TranspositionIdentities}}
Here we list a number of identities obeyed by the transposition brackets \eqref{TranspositionBrackets}. The following formula follows directly from \eqref{TranspositionBracketProperties1}:
\begin{equation} \label{TranspositionBracketProperties2}
\left\langle \Pi_s \gamma_\mu, \Gamma\right\rangle_r = \Pi_{-s}\left[\gamma_\mu,\Gamma\right]_r + \frac{s\,r}{2}\,\left[\gamma_4, \Gamma\right]_+ \gamma_\mu.
\end{equation}
Several particular cases of \eqref{TranspositionBracketProperties2}, used in the main text, are:
\begin{eqnarray}\label{TranspositionBracketProperties3}
&& \left\langle \Pi_\pm \gamma_\mu, \gamma_\nu \right\rangle_+ = 2\eta_{\mu\nu} \Pi_\mp \nonumber \\
&& \left\langle \Pi_\pm \gamma_\mu, \gamma_\nu \right\rangle_- = 2\gamma_{\mu\nu} \Pi_\mp \nonumber \\
&& \left\langle \Pi_\pm \gamma_\mu, \Pi_\pm \right\rangle_+ = \gamma_\mu \nonumber \\
&& \left\langle \Pi_\pm \gamma_\mu, \Pi_\mp \right\rangle_+ = 0 \nonumber \\
&& \left\langle \Pi_\pm \gamma_\mu, \Pi_\pm\right\rangle_- = \mp\gamma_4 \gamma_\mu \nonumber \\
&& \left\langle \Pi_\pm \gamma_\mu, \Pi_\mp\right\rangle_- = 0.
\end{eqnarray}

\bibliographystyle{nb}
\bibliography{refs}

\end{document}